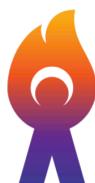

# POWERSHARE

# Mechanics

Draft 4.1

Beka Dalakishvili

Ana Mikatadze

Last Revision 15 February, 2019

Tbilisi, Georgia

# Table of Contents

# Introduction

Blockchain startups around the world are challenged by scalability and security issues of the underlying technology, real world application and end-user friendliness of their products, proper incentivization of the community and balanced valuation of their native currencies. Inability to overcome these challenges hinder broad adoption of blockchain technology, as well as its deployment for real-world use in general.

In this paper we present mechanics of Powershare - a gamified social network for charity crowdfunding fueled by public computing - and propose solutions to reach optimal balance between efficiency and resilience necessary for building sustainable, long-term future.

Powershare enables backers to financially support small personal causes without paying a penny from their own pockets but rather by sharing the idle processing potential (IPP) of their electronic devices. Donated processing power is utilized for public computing to support computationally heavy scientific researches of authorized research institutions that in return for the valuable resource maintain the decentralized open ledger of digital currency - FIRE. FIRE is the basic denomination unit and accounting currency within the Powershare ecosystem that enables monetization of users' computing power as well as value distribution within the network.

Accumulated computing resource donated by backers, along with the minted FIRE and every participant's non-monetary contribution of curating existing campaigns or creating new ones represent a common pool resource, hence, to avoid the "tragedy of the commons" and maximize network value, we exercise the eight governance principles offered by Elinor Ostrom (Ostrom, 2015), and design network architecture so as to allow high group formation capability, following the GFN law (Reed, 1999).

Major drawbacks of traditional charity crowdfunding platforms are (i) the narrowness of supporting audience, which is limited to medium to high income individuals. (ii) Extremely low retention rates - as little as 26% for the second donation, and only 1% for the fifth (Althoff, Leskovec, 2015) (iii) High platform and payment fees; and (Crowdfunding.com) (iv) high costs associated with promotion of the causes (Recomendes, 2015)

Our model has several major advantages over existing ones – firstly, backers have the opportunity to financially support charitable causes without monetary expenditure, moreover, directing the shared processing power towards scientific research creates additional layer of incentives. On top of social, psychological and emotional benefits derived from the act of giving, the backers donate through an engaging, gamified social network and are rewarded with digital tokens (FIRE) for supporting successful campaigns this on one hand increases retention rate of the backers and on the other simplifies asking for support. Powershare platform charges no platform or payment fees (in case of FIRE donation).

Above aforementioned advantages over traditional crowdfunding platforms, Powershare provides the ground for broad adoption of the blockchain technology, as users are not required to have any understanding of the technology in order to be part of the community. The process is demonstrated on the Alpha version of the platform currently displayed at www.powershare.fund. Where backers can become supporting heroes with just one click of a button, however the existing version, requires technological optimization.

Naturally, building such complex crowdsourcing system arises essential challenges to be resolved. This paper aims to propose

(i) Computationally executable mechanism design that matches the utility function of each network participant with that of the community. This is achieved by directly connecting fundraisers, backers and research institutions, enabling them to exchange "items of unequal value", that enables expanding the gain of each participant in this interaction (Diamond, 2010).

(ii) Governance framework of a self-organizing community where any user can obtain governing power through demonstrated contribution to the system in different ways. We propose to use gamification for increasing user engagement as well as for introducing our complex system architecture as a fun and exciting game to our users.

(iii) Multi-layer incentivization structure that maximizes value creation within the ecosystem and pushes aggregate utility of the network outwards. Supporting scientific research along with social and environmental causes increases emotional and social benefits derived, in addition to monetary rewards and exchange network incentivization.

(iv) Optimal scarce resource allocation model where campaigns voted by the community distribute it amongst themselves fairly.

(iv) technological configuration of the FIRE Consensus Protocol, a hybrid of Proof of Stake and Proof of Authority, which diminishes major problems of blockchain technology (such as scalability, security, unfair value distribution and conflict of interests between miners and users) and leverages its versatility.

# Background

## The economics of giving

Economic theories are mostly based on the assumption that all humans are self-interested in the sense that they always put their individual preferences in front of others, always trying to maximize their own *utility* by eliciting the most personal benefits, be it financial, emotional (ex.: appreciation from family, friends, and surrounding community), or social (ex.: social status, reputation) (Smith, 1962). However, when these theories are put to a test, the results repeatedly prove against this hypothesis (Güth et al. 1982), (Berg et al. 1995). Kolm and Jean Mercier Ythier offer a profoundly different view of society that should extend to the settings of mechanism design as well. In the first chapter of *The handbook of the economics of giving, altruism and reciprocity,* Serge-Christophe Kolm refers to the "un-selfish side" of humans:

"Altruism, giving and pro-social conduct, and reciprocity, are the basis of the existence and performance of societies, through their various occurrences… for remedying "failures" of markets and organizations (which they sometimes also create); and in charity and specific organizations." (Kolm, 2006)

Even though altruism is a common term for economists, empirical evidence refutes against some assumptions derived from it. By definition, altruism refers to the concept of giving without expecting something in return, and implies that a positive change in other's welfare positively changes the giver's utility as well (Andreoni, 2002). This phenomenon was thought to cause the crowding-out effect, which implies that if altruists' main concern is just the social welfare of others, then, when government increases donations to a charity, they should decrease their contributions accordingly. The results of the experiments however, have not always met the expectations: Brooks found no crowding-out effect in an experiment on public funding of symphony orchestra (Brooks, 1999), while on that of public radio funding he even identified crowding-in effect (Brooks, 2003), where the increase of the government's contributions increased that of the individuals' too.

Induced by such inconsistencies, Kingma suggested that the broad definition of charity as "aggregate contributions of individuals to social welfare or other organizations" does not differentiate them from one another while crowding-out effect actually varies through different types of charities (Kingma, 1989). The aforementioned different results of Brooks' experiments also pointed at what Andreoni would call "impure altruism", which suggests that on top of the overall level of charity, one's utility is affected by individual contribution as well. The satisfaction derived from the latter was labeled the warm-glow effect, or the joy of giving (Andreoni, 1989). Psychological benefits, one of the eight drivers of donation behavior, explains the warm glow effect as a "contribution to one's self-image", presenting one as a generous, empathic, and socially responsible (Bekkers, Wiepking, 2011).

Another driver for charitable giving is thought to be the awareness of need, which partially derives from subjects' direct exposure to certain troubles, and induces donation behavior merely because of the resemblance between one's own troubles and that of the others (Bekkers, 2008), (Burgoyne et al. 2005), (Olsen & Eidem, 2003). This leads to a very interesting phenomena, empathy.

Preston and De Waal offer a common definition of empathy as an understanding of somebody else's emotional state through the perception of their situation. Furthermore, they suggest that "an individual employs the brain's way of understanding the world to understand the internal states of others" through identification, emotional contagion, and cognitive empathy, implying that helping others is always in some way beneficial for the helper (Preston, De Waal, 2002). By investigating neuroeconomic evidence, Ernst Fehr and Klaus Schmidt suggest that what they call "other-regarding preferences", i.e. the preferences of an individual that are not derived from personal interests, can be better understood through empathic responses between individuals with different relationships. They bring forth evidence supported by brain imaging techniques such as Magnetic Resonance Imaging (MRI) that show how people empathize for peers they find to be fair, but not for those who have been unfair to them in previous interactions. Further experimental results on activations in the Striatum (main reward-related component in the midbrain) show that when punishing unfair players, activation in striatum is present. Even more, the subjects would punish unfair players even at their own expense. But the most important finding of neuroscience is that individuals gain higher satisfaction after reaching the goal through cooperation rather than in single-player games (Fehr, Schmidt, 2006).

In short, based on the neuroeconomic experiments, Ernst Fehr and Klaus Schmidt concluded that humans have a sense of "mutual cooperation and the punishment of norm violators", once again contradicting the self-interested hypothesis.

## Social Physics

Social physics is a relatively new discipline in science that, as described by Alex Pentland, a pioneer in this field, "uses game theory to mathematically examine the properties of human societies, such as comparing a society based on exchange networks with one based on markets" (Clippinger, Bollier, 2014).

By analyzing very large data sets that are far too complex for traditional processing software (hence the name Big Data) and by inheriting mathematical laws of biology, social physics offers a better understanding of human behavior (Socialphysics.org). The relationships humans hold between one another, and the ideas they exchange, forms and shapes the society as a whole. This corresponds to the theory of living systems, holonics, which represents the world as "nested hierarchies of subsystems", made up by individuals who cooperate and engage with each other. In holarchy, in contrary to hierarchy, individuals work together for a common purpose that reflects the social norms established through perpetual interaction of its peers while still maintaining their autonomy. Oversimplified market view, where individuals are seen as purely self-interested, failed to represent the inherent complexity of societies, however technological breakthrough provides powerful insights into how these ecosystems work. People follow and 'obey' the rules and norms around them, moreover, they communicate with different people and possess different, incomplete information about the circumstances, which basically implies that in reality societies should be seen as exchange networks, made up by specific exchange relationships, rather than as alien buyers and sellers in a perfect-information market setting. Stability, when viewed from this perspective, comes from the trust that is formed between the community members (Ulieru, 2014).

Natural selection unveils the secrets of sustaining such complex flow systems, where the optimum zone (a.k.a. the window of viability), for survival throughout evolution lies between efficiency, i.e. capacity of complex flow systems to process given volume, and resilience, i.e. the capacity of complex flow systems to survive. Within living systems, cooperation not only ensures trust, so crucial for stability of exchange networks, but moreover aids its positioning within the window of viability (Clippinger, Bollier, 2014).

The sense of cooperation also corresponds to the warm-glow effect, which, as suggested above, contributes to one's perception as a socially responsible being. However, this property is not universal. An experiment done on Kiva, an online-lending platform for low-income individuals, where the eight drivers of donation behavior were put to a test to reveal their significance, found that "People who feel that altruism is important for their reputation are more likely to report giving philanthropically." (Srkoc, Zarim, 2014)

These results extend to social learning - a process of learning from and inheriting the habits and beliefs of surrounding community, which depends heavily on idea flows - the spreading of ideas within the exchange network. Social learning is a natural process, which is stronger, safer, and more frequent than independent formation of individual preferences.

Moreover, solicitation appeared to be the most significant of the eight drivers on Kiva (Srkoc, Zarim, 2014). Many researches from past decades have shown quite strong correlation between solicitation and donation behavior, about 85% to 86% (Bryant et al. 2003), (Bekkers, 2005). However, studies suggest that the effectiveness of this mechanism depends on who is asking, and some conclude that backers are more likely to donate when they are asked by someone reputable, someone they know and have a relationship with (Sokolowski, 1996), (Bekkers, 2004).

The process of social learning incentivizes exchange networks more effectively than individual incentivization. In one of his presentations, Alex Pentland demonstrates an experiment where one group of individuals were rewarded when they became more social, and the other group was rewarded if their friend got more social. Given that by doing so the researchers created a topic of conversation and induced cooperation, the second group registered four to eight times more effective results per dollar. Interpreting this new approach of incentivization as "stretching the social fabric" corresponds to the importance of human interactions (Pentland, 2016).

Social Physics enables precise prediction of individuals' behavior as well as that of the community's as a whole through analyzing Big Data: "Because "idea flow" takes into account the variables of a social network structure, the strength of social influence between people, and individual susceptibilities to new ideas, it also serves another vital role: It gives a reliable, mathematical prediction of how changing any of these variables will change the performance of all the people in the network. Thus, the mathematical framework of idea flow allows us to tune social networks in order to make better decisions and achieve better results (Clippinger, Bollier, 2014)."

# POWERSHARE mechanism

Taking the game theoretic and microeconomic viewpoints into account, we create such initial conditions for network participants that lead to desired outcome of shifting the utility curve of the community outwards, which can only be achieved through cooperation. Vernon Smith and many others after him considered agents to be purely self-interested (Smith, 1962), however some empirical evidence shows that this assumption is not entirely true in practice, especially when predicting agents' behavior in such philanthropic setting as charity crowdfunding (Güth et al. 1982), (Berg et al. 1995). Hence, we broadly define the utility function of our users as one consisting of the agent's personal gain, their contribution, and overall level of donations, and we design the mechanics of the Powershare network accordingly. However, this definition will be formalized after deeper theoretical and experimental research, which remains future work. Existing model will serve as a determinant of the optimal outcomes more precisely, which is paramount in establishing the structure of incentives for each participant in a way that their utility function highly correlates with that of the community.

## Overview

Powershare is a charity crowdfunding platform that enables backers to financially support small personal causes by sharing the idle processing power of their devices. For doing so backers (and the campaigns they support) receive piece of block reward proportional to their contribution. The collected computing resource is directed towards scientific research of authorized research institutions, rather than towards maintaining the PoW algorithm as was the case in the earlier stages of the project. Currently, Alpha platform is retrievable at www.powershare.fund where donation of computing resource is achieved via browser-based mining of Monero (XMR) through scripts by Coinhive. Next milestone of platform development is to replace browser-based mining with public resource computing.

## Incentive Structure

Incentive Theory of motivation suggests that intrinsic incentives are longer lasting and do not satiate easily, since they correspond with one's inner desires and goals, such as self-realization, independence, and self-expression (Franzoi, 2003). Using such strong incentives for "stretching the social fabric" presumably will be more efficient and lead to greater performance, embracing collective charitable behavior, and satisfying inner needs of social activity.

As discussed above, mechanisms that encourage contributions from individuals to any type of charity include awareness of need, solicitation, altruism, reputation, and psychological benefits. Additionally, the campaigners' reputation as well as quality and popularity of the content positively influences perceived credibility, which in turn stimulates the backers' intention to donate (Bekkers, Wiepking, 2011).

Vernon Smith suggested that agents are purely self-interested, hence their utility solely depends on their personal preferences (Smith, 1962). In contrary, Ernst Fehr and Klaus Schmidt suggest that self-interest is conditional and occurs only in certain environments (Fehr, Schmidt, 2006). We strongly believe that in a setting of charity crowdfunding common welfare becomes a major variable of the individual utility function. Based on this assumption, agents' utility has multiple variables and is not limited to pure self-gain, i.e. does not consist of only self-benefiting actions. Therefore, Powershare incorporates a multi-layer incentive structure for all participants of the network:

*Backers* - gain social, psychological, emotional and monetary benefits from supporting charitable causes. Trivial financial rewards are applied through the feature of **mutual benefit** - meaning that the backers, when donating their idle CPU power, receive half of their contribution back, which they can either withdraw, make micro transactions on the platform, trade on an open market, lock to participate in governance of the network or donate the sum back to campaigns directly. Social, psychological and emotional benefits extend to additional layer as the processing power they donate not only helps the selected campaigns but also supports scientific research for the better of humankind. From game theoretic viewpoint, backers attain social impact and generate monetary reward only through cooperation, since crowdsourcing in itself implies that all work together for common goal. The non-linear distribution algorithm of reward allocation ensures every backer has the ultimate motivation to seek to support campaigns that have higher probability to be funded, which leads to higher cooperation and group forming within the community as well as to higher quality of the campaigns.

*Campaigners (a.k.a. Orators)* - request funding for their small personal goals or on somebody else's behalf. Their major motivation is monetary, hence the nature of the platform puts them in a competitive game. As Liu et al. imply, the success of campaigns in a micro charity setting derives from their perceived credibility and the degree to which they can induce empathy within the supporting community (Liu et al. 2017). For that, reputation of the initiator as well as the quality and popularity of the campaign are paramount. Measurement of campaigner's reputation and popularity of the cause are achieved through gamification (ex.: title of the campaigner derived from his/her contribution, reputation bar - determined through multiple variables, campaign rating by the community). Gamification framework is further discussed below.

To summarize, the only way Orators can defect is to provide fake campaigns, however given the gamified design of the platform ensures honest participants have more chances to succeed, moreover neuroscientific evidence suggests that people have a sense of punishment of norm violators (Fehr, Schmidt, 2006), making the optimal decision of each participant to act honestly.

Presumably, incremental incentive for the Orators is that by providing high quality, credible and popular campaigns, they assist scientific research fueled by processing power they received.

*Senators* - the authorities of the network are research institutions that require processing Big Data for the better of humankind (Anderson, 2004). Senators' utility is proportional to the aggregated processing power of the network. In return for the computational resource they operate the underlying blockchain. Having no competition on the approval of a new block and moreover, having no share in the block reward puts them in a cooperative game. Their cooperation extends to splitting the computing resource amongst themselves equally[1].

---

[1] Under Consideration

# Flexibility of the network

The properties of self-organization are paramount for not only long-term sustainability of the network, but also for reaching new heights in value creation. To allow an open network to reach its highest value of $n^2$, where n is the number of network participants, the architecture itself must enable digital governing, and provide sufficient tools for self-reflection and "law enforcement" on individual levels. Such approach replaces the well-known broadcasting model of sharing the created content with "peer-to-peer transactions" where every user is participating in content evaluation (Bollier, Clippinger, 2013). To establish such network architecture suggested by David Reed, who denominates this model as the Group Forming Network (GFN), the Powershare ecosystem enables community governance in several ways, mainly through the use of gamification and voting:

Powershare uses the techniques of gamification for implementing its community-driven governance to achieve a higher level of participation and simplicity. Each and every participant of the network is awarded different titles (such as "Tourist", "Craftsman", "Citizen", "Hero" etc.) and levels within those titles based on their contribution to the community as a whole (variables include amount / CPU donated, quality of initiated and supported campaigns, participation in voting, sharing content on social media and bringing active friends on the platform). Apart from badges, leaderboards and achievements, the game design is being developed under the Octalyses framework (Chou, 2015), which ensures higher levels of interaction and user retention rates through provoking sense of ownership and accomplishment among many others. Notably, in order for a user to become an Orator they have to advance beyond entry level (Level 1 Tourist). This creates the minimum threshold of contribution (and verification) before opening a campaign and is useful for tracking campaigners' credibility. The higher the level/title of a user, the higher quality campaign they are able to orate, as users with higher levels unlock different tools and features to use their creativity to its maximum - they become able to design their profile, avatar, campaign pages and layout and provide more texts, photos and videos. Furthermore, they will be able to interact with the community, receive voting rights, statistics about supporters etc. Different techniques will be used to incentivize giving as well. For example, those who donated the most will be put out for others to see, and will receive special badges and personalized letters from those helped. Moreover, Powershare will offer affiliate programs, meaning if a user brings a friend on the platform who gets successfully active, the user will get rewarded for it.

Additionally, since campaigns get funded based on the current trends within the community - which most likely will be quite differentiated at any given time period - backers will form alliances and collectively support campaigns of backgrounds and goals matching that of their owns'. To allow efficient spreading of ideas and support group formation of like-minded participants, the platform is designed to be a social network for donation-based crowdfunding, where users can see each other's profiles, past activity, aggregate contribution to the network, reputation. Plus they can follow and mimic the users to ease group forming and easily detect campaigns they might be interested in via push notifications and other features.

Another major tool to increase the ability of the community to self-organize and govern on their own is voting. Users with the Citizen rank or higher have a vote in resource allocation - a process of distributing failed campaigns' funds amongst the active ones. To achieve the rank of a Citizen, one must lock some amount of FIRE (TBD) on their accounts, go through a KYC process etc. Besides network participants are eligible to vote on addition of a new authority[2], where proposition is made by JSC Powershare Foundation.

## Tokenomics

Another significant gear in the mechanics of the Powershare network is the monetary value of the digital currency - FIRE, which is the basic accounting unit within the ecosystem and its monetary properties are reinforced with underlying blockchain. We consider VEN as the benchmark digital currency and reframe its Four Core Attributes for FIRE. The goal is not to mimic VEN but rather inherit its advantages and merge them with the characteristics of decentralized currencies.

*Resilience* of FIRE price towards extreme market conditions is paramount, as the network is self-organizing, self-reflexive and self-governing fixing the FIRE price to a stable exchange rate will restrict it from reflecting the up to date state of the ecosystem as well as expectations of the community. Rigid valuation of FIRE will decrease its efficiency, hence flexibility is seen as key to sustaining the network, while defending the price against extreme fluctuations is an absolute necessity. The valuation of FIRE naturally depends on fundamental economic principles of supply and demand, as well as total outstanding supply of the coins and usage of the reserves possessed by Power Share Foundation.

*Globality* of FIRE derives from the nature and user-base of the Powershare platform. As anybody around the globe with access to internet is a potential user, outstanding supply of FIRE will be geographically distributed. The major challenge of the foundation is to make it accessible worldwide and to ensure fast and easy exchange of FIRE to fiat currencies.

*Security* is one of the major reasons behind the rise of cryptocurrencies. Cryptography behind them as well as their decentralized nature are the keys to its security and immutability. However, 51% attacks and double-spending problems have been witnessed in multiple networks (Hertig, 2018). Also the conflict of interests between miners and users is a major problem. To tackle this drawbacks, FIRE is a permissioned blockchain where only authorized full nodes are eligible to participate in the consensus (further details in The Fire Consensus Protocol), enabling Powershare to utilize the versatility of the technology while maintaining its credibility and security.

*Social Impact* is deemed as the fourth core attribute of FIRE, which derives from the very nature of the currency and the ecosystem that it fuels. FIRE can only be minted through supporting charitable causes and by providing computing power to scientific research of authorized institutions. One of the greatest challenges posed by digital, decentralized currencies is the fact that they have vague intrinsic value, which is usually defined as the "security of the network", "utility of the token" etc. We define the intrinsic value of FIRE as the social impact derived from donation-based crowdfunding and contribution to the welfare of humankind of the scientific research that utilizes public resource computing through the Powershare platform. Nevertheless, the valuation of the humankind welfare is still quite ambiguous, hence the minimum market valuation of all outstanding FIRE would be the price of network aggregate processing power on the supercomputing market.

FIRE price, apart from its intrinsic value described earlier is affected by fundamental economic principles of supply and demand, which derive from the utility and usability of the currency. Supply increases with every new block, however it is added to the circulation only once campaigns reach the funding goals. Funds aggregated on campaign accounts that fail to reach the funding goals are redistributed amongst the community, while the same amount is pulled out of the circulating supply making it a zero sum game. Circulating supply is further decreased through the holdings of the Authorities of the network which are used as a financial penalty if they behave in a dishonest manner and try to validate invalid blocks, while locked amounts on the backers accounts enable them to vote in the ballots designed for self-governance of the network. FIRE can further be used for payments across the network, micro transactions on the platform and direct donations to campaigns.

---

[2] Under discussion

Speculation is not taken into account as it affects both sides of the equation, moreover we design FIRE to be a "yawn for speculators... and boon for common people" (Stalnaker, 2014).

FIRE price is further balanced by the foundation reserves, which are used to defend the price against malicious traders and extreme market conditions. JSC Power Share Foundation holds 10-20% of all outstanding FIRE at all times and regulates the valuation in a semi-floating manner, i.e. the foundation will act as the largest market participant to balance the exchange rates.

Further balancing of FIRE valuation is achieved through smooth emission of the currency described and formalized in the Reduction of circulating supply and Appendix #1.

## The Blockchain

Currently Powershare has developed an alpha platform that can be retrieved at www.powershare.fund. The platform operates using the scripts by Coinhive, which enables backers to donate their processing power towards charitable causes. The accumulated processing power is than pooled and used to sustain the Egalitarian Proof-of-Work algorithm of a CryptoNote currency - Monero (XMR). However, the value created during the process is negligible as: (i) it does not significantly increase the total hash rate and hence the security of the Monero network, (ii) the current market conditions result in smaller dollar value created during the process and (iii) Coinhive charges 30% fee on all minted currency. Naturally, the first step towards improving the performance of the model was seen in three major steps of replicating Coinhive scripts, providing pool management for collected processing power, and creating a CryptoNote based currency – FIRE, that would enable higher efficiency of the existing model. The practical implementation and reasons of choosing CryptoNote are further detailed in appendix #1. However, CryptoNote currencies do not provide the ground to build smart contract and governance layers on top of the aforementioned technology and create significant security risks (Sun et al. 2017). Hence, we present FIRE consensus protocol described hereafter.

### The FIRE Consensus Protocol (FCR)

FCR is being designed to fit the complex structure and value flow of the network. Accumulated resource represents a common pool for everyone to benefit from, and to enable its fair distribution over a long period of time, avoiding the tragedy of the commons, FCR leverages the versatility of blockchain technology while overcoming the shortfalls of absolute decentralization. The protocol is a modification of Proof-of-Authority (POA) with extensions from the logic implemented through Proof-of-Stake (PoS) - Ethereum Casper in particular. FIRE, being a permissioned blockchain, only has a small number (TBD) of trusted full nodes that maintain the network on behalf of a purposefully designed two-tier incentive structure while still enabling community driven governance and content evaluation on the platform layer.

These trusted nodes (a.k.a. the Senators) will be the respected research institutions, that utilize the processing potential of the network, for scientific research, for the better of the humankind, rather than waste the precious resource on maintaining the PoW protocol. In exchange, Senators are obliged to maintain the blockchain and lock a fair amount (TBD) of FIRE, which will be used as a penalty in case of dishonest actions of the Senator, additionally serving as a means of reducing circulating supply of FIRE.

## Blockchain maintenance

The Senators stake their authority and FIRE collateral while participating in the consensus that provides encrypted append only open ledger of the distribution of the value within the network. Block reward is allocated to campaigns and backers based on the shared power through and by their accounts. Consensus requires 2/3 of the votes of the authorities[3]. Invalid block validation results in loss of the staked amount as well as authority of the Senator. In contrast with the PoW algorithm, FCR requires negligible processing power for maintaining the Blockchain.

## Resource allocation

The building blocks of Powershare ecosystem ensure an increase in all participants' utility within the network. As described earlier, the main value of the Authorities is the extracted amount of Idle Processing Potential (IPP) from participants' devices. On the other hand, campaign initiators are given the opportunity to fulfill their personal goals and dreams. Backers are socially incentivized to help those in need, along with supporting research for the better of humankind, and represent the main driving force of the network. They select charitable causes they are willing to support financially and do so proportionally to how much power they share. Powershare web platform provides the frontline incentives for the community, as the value created on behalf of the campaigns is aggregated on a smart contract (Contr), with built-in nonlinear resource allocation algorithm. Contract (Contr) distributes the accumulated amount with the logic outlined in Figure 1:

---

[3] Under Discussion

```
1: def funding(campaign, backers):
2:     for backer in backers:
3:         backer.relative_contribution = backer.cpu_donation/campaign.cpu_donations
4:
5:     if campaign.balance == campaign.required_amount:
6:         campaign.reward(campaign.direct_donations + campaign.cpu_donations/2)
7:         for backer in backers:
8:             backer.reward(backer.relative_contribution * campaign.cpu_donations)
9:         surplus_allocation = 0
10:        circulation_reduction = 0
11:
12:    elif campaign.balance > campaign.required_amount/2:
13:        campaign.reward(campaign.direct_donations + campaign.cpu_donations/4)
14:        for backer in backers:
15:            backer.reward(backer.relative_contribution * campaign.cpu_donations/4)
16:        surplus_allocation = campaign.cpu_donations/4
17:        circulation_reduction = campaign.cpu_donations/4
18:
19:    else:
20:        campaign.reward(campaign.direct_donations)
21:        #No reward for backers
22:        surplus_allocation = campaign.cpu_donations/2
23:        circulation_reduction = campaign.cpu_donations/2
24:
25:    return surplus_allocation, circulation_reduction
```

Figure 1: *the provided code represents the logic of the non-linear reward allocation to the campaigns and backers, as well as the logic behind reduction of the circulating supply and accumulation of the surplus funds of the network to be allocated to campaigns through voting.*

As the network benefits from utilizing public processing power, community in turn receives a piece of the new mint of every block. More precisely, amount to be aggregated on each campaign contract equals weighted average processing power donated to the campaigns multiplied by the mint of a new block:

fund_contr = campaign.cpu_donations / network.cpu_donations * mint

Notably, the authorities do not receive pieces of the block reward. In contrary, they stake their belongings on each block outcome, which is used to reduce circulating supply of FIRE if the authorities validate an invalid block, acting as a financial penalty mechanism for defecting.

Outstanding circulating supply is further affected by the locked coins of the Citizens, who are in turn awarded a vote in the surplus allocation and authority elections[4].

## Surplus allocation

Surplus is deemed as half (or quarter) of the FIRE balance on the contracts that could not reach the fundraising goal by deadline, based on the algorithm provided below. If the contract deadline expires and the balance of the contract is less than the hard cap (soft cap) 0.5 (0.25) of total funds, they are aggregated on another contract (Contr_surplus_allocation), which is than distributed amongst the eligible campaigns based on the voting within the community.

Firstly, the number of campaigns ($n$) to be funded is calculated given the monetary value to be distributed amongst them and aggregate remaining amount required by campaigns to collect requested funding,

$n$ is the largest number where:

$\sum_{i=1}^{n} a_i$ <= surplus_allocation * sufficiency_coefficient

(sufficiency_coefficient) on the other hand is a factor by which the surplus allocation is sufficient to cover aggregate remaining deficit to full funding of the $n$ campaigns.

Once the number of campaigns that will share the common pool resource is determined, in order to fairly split the funds amongst the selected campaigns each of them should receive equal amount, however there will presumably be campaigns that require less than average amount to reach their funding goals. These campaigns are funded first, which increases the average amount to be distributed amongst remaining campaigns. In this way, those requiring less get full funding, while those requiring more do get more. The implementation of the described logic is provided in below:

---

[4] Under Discussion

```
1: def surplus_distribution(surplus_allocation, campaigns):
2:     for campaign in campaigns:
3:         campaign.left_to_fill = campaign.required_amount - campaign.balance
4:
5:     while True:
6:         avg_amount = surplus_allocation/len(campaigns)
7:         for campaign in campaigns:
8:             if campaign.left_to_fill <= avg_amount:
9:                 campaign.reward(campaign.left_to_fill)
10:                surplus_allocation -= campaign.left_to_fill
11:                campaigns.remove(campaign)
12:                end_cycle = False
13:
14:         if end_cycle:
15:             break
16:
17:     for campaign in campaigns:
18:         campaign.reward(avg_amount)
```

Figure 2: *The provided code represents the logic of surplus allocation, where funds attributed to failed campaigns is distributed amongst the campaigns voted by the community*

### Reduction of circulating supply

As shown in the algorithm (surplus_allocation = circulation_reduction), which is linked to the decrease of total outstanding FIRE for the moment. For doing so, typically a burn mechanism is used. Despite its positive effect on the tokenomics, this method implies that portion of the value created within the network is being destroyed. In order to maintain the total value created within the network, we propose "Vaporization" - where the tokens are added back to the mintable amount of coins. Which will result in a dynamic block reward amount as well as smoother emission and prevention of total supply extortion.

The block reward is defined as:

$$BaseReward = (MSupply - A) >> 23 ,$$

Where A is the amount of minted coins,

Proposed algorithm is to introduce another variable of R into the equation, which increases the block reward based on the amount of {circulation_reduction}, hence the block reward formula will change accordingly:

BaseReward = (MSupply – (A-R))  >> 23

## Voting

There are three major votes within the network:

1. The consensus - transformation of the blockchain into a new state is voted amongst the Senators, consensus requires majority of ⅔ of total votes.

2. Elections[5] - Citizens vote on the addition of new authorities with a majority of ¾ of the vote.

3. Surplus fund allocation - Citizens vote on fund distribution amongst the campaigns, the $n$ number of campaigns with the most votes get funded from the common pool, with the logic outlined in previous section

Naturally, voting, being one of the key elements in the proposed decentralized governance structure, needs to be anonymous at least before the ballot is over. This significantly reduces security risks as well as improves economic soundness of the network.

## The building platform of FIRE consensus algorithm

The most widespread solution for smart contract implementation, necessary for the governance structure described above, is Ethereum. However, limitations of Ethereum might impede implementation of the FIRE Consensus Protocol.

1. *Privacy* - in Ethereum users are pseudonymous rather than anonymous, and a number of studies show that "pseudonymity provides only weak privacy protection. Moreover, *contract state and user input must be public* in order for miners to verify correct computation (Cox, 2018)." Hence, it is virtually impossible to conduct a private ballot outlined in the previous section.

2. *Scalability* - the mechanism design of the Powershare network requires complex calculations, in Ethereum these calculations are conducted on-chain, meaning that transition of the contracts, from one state to another, might associate with lack of timeliness.

3. *Transition costs* - "in August 2017, it cost $26.55 to add 2 numbers together one million times in an Ethereum smart contract" (Cheng et al. 2018), computation of the complex algorithms described in this document will inevitably result in higher Gas price paid for each state transition.

---

[5] Under Consideration

The proposed solution is that of EKIDEN that "leverages a novel architecture that separates consensus from execution, enabling efficient TEE-backed confidentiality-preserving smart contracts and high scalability" (Cheng et al. 2018) as well as reduces transaction costs. EKIDEN however is at the initial stage of development, but if executed as planned, it is closest to the requirements posed by the FIRE consensus protocol.

We believe that Blockchain is the missing piece in the technological revolution we are now witnessing, since it greatly contributes to a more distributed and fair allocation of power and wealth while maintaining immutability and security. As such, defining fairness is transformed into collective effort, enabling adjustment to individual circumstances as never before.

# Related Works

Extraction of Idle Processing Potential (IPP) of consumer hardware has been a challenge tackled by different projects. However, those solutions are fragmented by industry, geography, time and context.

## Browser-based mining

Scripts for browser based mining were first introduced by bitcoinplus.com in 2011, followed by TidBit in 2013. After both of those projects failed due to market conditions and legal issues (Cox, 2018), Coinhive was introduced in September 2017 (Symantec, 1982). Coinhive scripts are being utilized by the alpha platform of Powershare (at www.powershare.fund), as well as the prototype deployed in February 2017, and enables the model of CPU donation through a web browser. Similar scripts were designed by hackers that compromised a popular plugin used by thousands of websites, tweaking it to inject code which caused visitors' browsers to generate digital coins on the hackers' behalf. This practice is known under the name of cryptojacking, which has demonstrated an 8500% growth during 2017 (Saad, 2018).

Browser-based mining has been used for fundraising purposes by UNICEF Australia, through the web page www.thehopepage.org, which is backed by over 25k users since establishment in April 2017. The data from (Mohaisen et al., 2018) suggests, that the amount raised should be negligible, taking into account the inefficiency of mining Monero (XMR) with a personal computer, market conditions and the number of supporters. The major challenge in adoption of browser-based mining is the negligible reward for user's processing power resulting in negative profitability of the process.

## Berkeley University of California

### BOINC

A more effective way to utilize the CPU of consumer appliances was presented by Berkeley Open Infrastructure for Network Computing (BOINC) in the early 2000's. "BOINC *is a software system that makes it easy for scientists to create and operate public-resource computing projects"* (Anderson, 2004)*. In their research paper,* public-resource computing is seen as an alternative to maintaining supercomputer centers and dedicated machine rooms, which dramatically reduces the set-up and processing costs for scientific research. Moreover, the potential of consumer hardware in terms of floating point operations per second (FLOP/s) is significantly higher than that of hashes per second (h/s). The major challenge for BOINC projects is to adequately reward backers, in order to increase their incentives for the community as well as improve ease of user interaction - overall. As of December 2, 2018 BOINC was able to extract 27 petaFLOP/s from the global IPP, which is relatively the same power as world's 9th biggest supercomputer (Titan) has (Mordor Intelligence, 2017).

***World Community Grid*** - a project coordinated by IBM, was launched in November 2004 and "enables anyone with a computer, smartphone or tablet to donate their unused computing power to advance cutting-edge scientific research on topics related to health, poverty and sustainability" (World Community

Grid, 2014). World Community Grid has 47,599 active users contributing 792,885 TeraFLOP/s as of December 2, even though WCG software is available on Windows, IOS, LINUX, Android the project still utilizes a trivial portion of the market potential (Boincstats, 2004). One of the probable reasons of low efficiency is mediocre incentive structure and ease of access, as all BOINC projects require the supporters to install software.

"**Gridcoin** is a blockchain-based distributed computing network powered by the idle processing potential of existing hardware." According to the whitepaper, Gridcoin targets the IoT industry, as any device housing a processor can be part of the network (Gridcoin Whitepaper, 2017). As of December 2, 2018, Gridcoin is the largest contributing team to BOINC, exceeding 25% of the total collected processing power with 3,019 active users (Boincstats, 2004). The incentives of Gridcoin community is proportional to the provided processing power of the supporters. However, Gridcoin token price has dropped by 97.3% from the peak price of 18.8 cents on January 7, 2018 and equals 0.4991 at the moment of writing (Coinmarketcap). This numbers lead to a conclusion that the major motivators of Gridcoin's existing users are the social and emotional benefits rather than monetary.

## Stanford University

### Folding@home

Folding@home (foldingathome.org) is a distributed computing project which enables the donation of CPU/GPU via a simple software API for simulating protein folding. The scientific research is directed towards diseases like cancer, Alzheimers, Huntingtons, etc. The project is founded by Pande Lab, which is part of the Departments of Chemistry and of Structural Biology, Stanford University and Stanford University Medical Center. With 135 petaFLOPS Folding@home is the largest network of public resource computing, running on contributions from almost two million people since 2000 (Stats.Foldingathome).

### CureCoin

Curecoin (curecoin.net), launched in 2014, is a Blockchain-based currency running on PoW algorithm, which maintains the blockchain with ASIC miners while enabling users to donate their GPU/CPU to Folding@home project. Curecoin produces 45 PetaFlops, being first in rank with more than 7 thousand members. For incentivizing contributors, it uses an automated distribution system from cryptobullionpools.com, and a Proof of Stake based system for incentivizing ASIC miners who ensure the security of the underlying blockchain. Although, currently the project is planning on a hard fork to move from PoW to PoS, in order to remove" the energy-hungry ASIC variable from the system."

# Conclusion

Powershare's uniqueness derives from the multi-layer incentive structure. The whole ecosystem is designed to be a self-organizing network of exchanges where every participant has a mission to accomplish, value to create for the better of the network and in return enjoy tailor-made incentives for doing so. The backers - the engine of the ecosystem - are not only motivated by ambiguous scientific research that utilizes Idle Processing Potential of their IoT devices, or monetary rewards for sharing their computing power. Their motivation rather derives from the social impact they make and the social, emotional and psychological benefits they gain by supporting charitable causes at virtually no cost. The Orators - not only raise funds to realize their campaigns but contribute to the scientific research as well. The rules of the platform and the competition amongst them ensures high quality content. And last but definitely not the least, the Senators enjoy valuable processing resource in return for maintaining the PoA blockchain algorithm and save the enormous costs of purchasing, setting up, maintaining and running supercomputers and dedicated machine rooms.

Community is a critical gear in the holonic Powershare system, since cooperation and group forming are the only ways to achieve common goals, govern the network and ensure long term sustainability.

Notably, UI/UX alongside gamification are paramount to engaging the users and increasing retention rates. Ease of use and the fun of doing so are crucial to grow the community and convert basic / non-tech savvy contributors to members of the community.

Another major advantage of Powershare is its technological solution, trusted nodes eliminate the difficulties posed by absolute decentralization while still maintaining the versatility of the blockchain. Solid economic model and utility behind FIRE, as well as its intrinsic value, ensures further viability of the open sector charity crowdfunding model.

# Future work

1. Quantification and implementation of logics described in this paper
2. Practical implementation of FIRE consensus protocol
3. Formalization of utility functions of each network participant as well as aggregate utility of the community.
4. Aggregation and analyses of empirical data on how (whether) the behaviour of agents will change in a charity crowdfunding setting when backers are not required to contribute financial resources.
5. Determination of optimal solution of circulating supply reduction.
6. Formalization of voting algorithm as well as specification of where ballots should be applied.
7. Development of a formal process of Authority assignment
8. Finalization of the Game design, research of further gamification techniques.
9. Inheriting Elinor Ostrom's eight design principles for commons pool resource management.
10. Further testing and upgrading the alpha platform

# Appendix #1- CryptoNote Implementation

FIRE was being designed as a CryptoNote currency in the beginning stages of POWERSHARE project development. This appendix illustrates the work that has been done and explains the reasons behind pivoting to other, more sophisticated option as described in the Fire Consensus Protocol.

The main reason behind replicating a CryptoNote currency is that replacing Monero (XMR) with FIRE on the alpha platform is easily executable. Moreover, there are several important characteristics of CryptoNote that create a sound basis to build upon:

*Egalitarian Philosophy* is fundamental to the Powershare project. Technically speaking, the value created by home appliances can only be exploited in an ecosystem with no competition from special purpose hardware. This idea is one of the major drivers of the CryptoNote protocol as well. As the Whitepaper - CryptoNote 2.0 by Nicolas van Saberhagen [35] states: "Our primary goal is to close the gap between CPU (majority) and GPU/FPGA/ASIC (minority) miners. It is appropriate that some users can have a certain advantage over others, but their investments should grow at least linearly with the power." (Saberhagen, 2013)

*Flexibility* is paramount for reaching the huge number of $n^2$ of value creation in Group Forming Networks (Clippinger, Bollier, 2014). The community's property of self-organization is an absolute requirement for building a sustainable ecosystem. In CryptoNote difficulty target, block size and transaction fees are dynamically adjusted by the algorithm, taking into account the current state of the network.

*Privacy* is essential to the idea of blockchain itself and is not limited to untraceable transactions. In CryptoNote this is achieved through the ring signatures, which on one hand enables private transactions, but on the other hand constraints building additional layers on top of the underlying blockchain (Sun et al. 2017) (for example creation of smart contracts), which is a major concern for further development.

One of the extraordinary features of CryptoNote is that it defines the optimal quantity of total supply as

$$MSupply = 2^{64} - 1$$

Which equals to 18446744073709551615 atomic units. "This is a natural restriction based only on implementation limits, not on intuition such as "N coins ought to be enough for anybody" [35]". (Saberhagen, 2013).

The total supply of coins translates into C code as:

```
const uint64_t MONEY_SUPPLY = (uint64_t)(-1);
```

Furthermore, the economic implication of the project, creates an intuitive currency supply restriction, which leads to FIRE having 10 decimal points

```
constsize_t CRYPTONOTE_DISPLAY_DECIMAL_POINT = 10;
```

Resulting in total supply of 1,844,674,407.

20% of total coin supply is pre-mined, 60% of which is being used to fund the development of the project, 12% is attributed (and locked) to the developers, the foundation and the team each get 10%, 5% is reserved for the community and the remainder (3%) for the advisors. Given the initial supply of 368,934,881, difficulty target of 10 seconds (const uint64_t DIFFICULTY_TARGET = 10) and emission speed factor of 23 (const unsigned EMISSION_SPEED_FACTOR = 23), the chart below, visually represents emission curve:

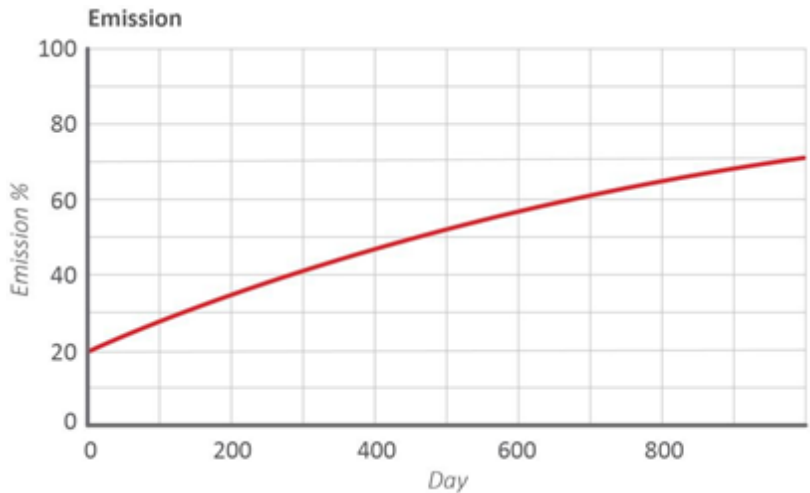

Graph below also illustrates the block reward curve for a CryptoNote currency with the above described settings:

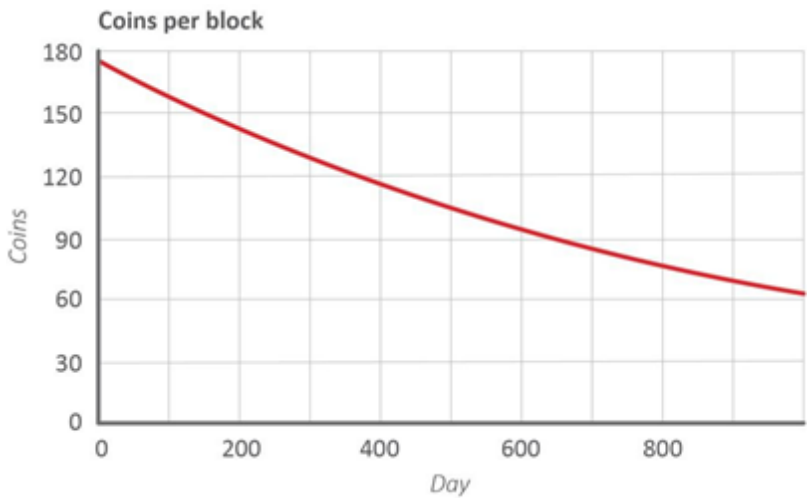

(Charts by Forknote.net)

Important variable that affects the block reward is block size., CryptoNote balances the size of each block based on the optimum of the network's current state. Which is defined as the median of the last N blocks, and larger block sizes will accrue excess size penalties. If M is deemed the median size of the last N blocks, the maximum acceptable size of a new block equals to

$2 * Mn$ this allows the block size to grow gradually and averts blockchain from flooding. Miners are being penalized for validating larger blocks. The penalty function is provided below:

$$NewReward = BaseReward \cdot \left(\frac{BlkSize}{M_N} - 1\right)^2$$

This translates in C code as:

```
bool    Currency::getBlockReward(size_t    medianSize,    size_t    currentBlockSize,    uint64_t
alreadyGeneratedCoins, uint64_t fee, uint64_t& reward, int64_t& emissionChange) const
```

Despite the above-mentioned benefits and practical implementation of a CryptoNote currency, capabilities of this protocol do not expand beyond monetary transactions and do not enable us to build smart contract layer on top of it, which is crucial for the project implementation. Moreover, past experience of mining industry has proven CPUs to be least efficient hardware for sustaining an open, decentralized and secure PoW algorithm leaving the network vulnerable to malicious actors (special purpose hardware).